\documentclass{iopart}

\pdfoutput=1

\usepackage{graphicx}
\usepackage{amssymb}
\newcommand{\hetArr}{\rightarrow}
\newcommand{\ie}{{i.e.}}        
\usepackage{hyperref}

\usepackage[sort,compress]{cite}

\graphicspath{{figs/}}

\begin{document}

\title[Relativistic solitary waves in plasmas]{Relativistic solitary waves modulating long laser pulses in plasmas}

\author{ G. S\'anchez-Arriaga, E. Siminos and E. Lefebvre}
\address{CEA, DAM, DIF, 91297 Arpajon, France}

\ead{erik.lefebvre@cea.fr}

\begin{abstract}

This article discusses the existence of solitary electromagnetic waves trapped in a self-generated Langmuir wave and  embedded in  an infinitely long circularly polarized electromagnetic wave propagating through a plasma. From the mathematical point of view they are exact solutions of the 1-dimensional relativistic cold fluid plasma model with nonvanishing boundary conditions. Under the assumption of traveling wave solutions with velocity $V$ and vector potential frequency $\omega$, the fluid model is reduced to a Hamiltonian system. The  solitary waves are  homoclinic (grey solitons) or heteroclinic (dark solitons) orbits to fixed points. By using a dynamical systems description of the Hamiltonian system and a spectral method, we identify  a great variety of solitary waves, including asymmetric ones, discuss their disappearance for certain parameter values, and classify them according to: (i) grey or dark character, (ii) the number of humps of the vector potential envelope and (iii) their symmetries. 
The solutions come in continuous families in the parametric $V-\omega$ plane and extend up to velocities that approach the speed of light. The stability of certain types of grey solitary waves is investigated with the aid of particle-in-cell simulations that demonstrate  their  propagation for a few tens of the inverse of the plasma frequency.

\end{abstract}

\pacs{52.35.Sb, 52.38.Kd}

\section{Introduction}

The excitation of long-lived solitary waves during the interaction of high-intensity laser pulses with plasmas is a topic with  applications and of theoretical interest. As multi-dimensional particle in cell (PIC) simulations have shown, these waves form  behind the laser pulse and they consist of electron density depressions with a trapped intense electromagnetic field oscillating at a frequency well below the laser frequency~\cite{Bulanov99,Sentoku99,Naumova01a,Naumova01b,Esirkepov02}. Solitary waves can  propagate towards the plasma-vacuum interface where the stored electromagnetic energy is radiated away in the form of low-frequency electromagnetic bursts~\cite{Sentoku99}, a process recently detected in the laboratory~\cite{Kando09}. Bright spots in optical plasma images with the same polarization as the laser pulse have been attributed to the formation of such solitons~\cite{Pirozhkov07}. These waves can evolve to a state named postsoliton, that has also been  observed in the laboratory with proton imaging techniques~\cite{Borghesi02a,Borghesi02b,Sarri10,Romagnani10}. 

The propagation of electromagnetic solitary waves has been intensively investigated within the cold, relativistic, one-dimensional fluid approximation \cite{Kozlov79,Kaw92,Kuehl93,Esirkepov98,Farina00,Farina01a,Poornakala02, Farina02,Lontano03,Farina05,Borhanian09}. Restricting attention to circularly polarized travelling wave solutions, with velocity $V$ and frequency $\omega$, yields a pair of second order differential equations that governs the dynamics of the electrostatic potential $\phi$ and the amplitude of the vector potential $a$. Hence, the system describes a Langmuir and an electromagnetic wave coupled by the nonlinear terms arising from the perturbation of the density and the relativistic mass. It admits solitary waves with vanishing (VBC)  and nonvanishing (NVBC) boundary conditions and, even though the governing equations are not completely integrable, they are commonly referred to as solitons.   

For VBC, $a\rightarrow 0$ and  $\phi\rightarrow 0$  as $x\rightarrow \pm\infty$, a soliton is interpreted as a light wave which is trapped in a self-generated plasma wave \cite{Kaw92}.  They  have been classified according to the number of zeros of the vector potential profile, $p$, see Ref.~\cite{Farina01a}. They are commonly referred to as bright solitons and their stability has been  analyzed too \cite{Naumova01b,Hadzievski02,Poornakala02,Poornakala02b,Lontano03,Mancic06,Lehmann06,Saxena07,Lehmann08}. In \cite{Farina01a}, the existence of such solutions is discussed in the parametric $V-\omega$ plane and $p=0,1,2...$ families of solitons are identified. Some of these families end at certain velocity values where the ion density profile shows a cusp at the center of the soliton. The soliton breaking  has been proposed as a mechanism for particle acceleration in high-intensity laser plasma interaction \cite{Farina01a} and it was observed in PIC simulations for solitons with $V=0$ and overcritical amplitude \cite{Esirkepov98}. Solitons with VBC in warm  \cite{Lontano03} and magnetized plasmas \cite{Farina00,Borhanian09} have been studied too.  

On the other hand, three different solutions with NVBC are possible \cite{Farina02,Farina05}: (i) grey solitons ($a\rightarrow a_0$ and  $\phi\rightarrow 0$  as $x\rightarrow\pm\infty$), (ii) dark solitons ($a\rightarrow \pm a_0$ and  $\phi\rightarrow 0$  as $x\rightarrow\pm\infty$) and (iii) shock waves ($a\rightarrow a_0$ and  $\phi\rightarrow 0$  as $x\rightarrow-\infty$ and $a\rightarrow 0$ and  $\phi\rightarrow \phi_0$  as $x\rightarrow +\infty$). The asymptotic values $a_0$ and $\phi_0$ are related to the two parameters $V$ and $\omega$. The branch of shock waves splits the $V-\omega$ plane in two different regions where  either dark or grey solitons exist. Branches of solutions are found and they break down at increasing $a_0$ due to  divergence of the electron density  \cite{Farina02,Farina05}.

As we will see, the grey and dark solitons with NVBC can be interpreted as a localized modulation in a long circularly polarized electromagnetic wave coupled with a plasma wave. Even though the circularly polarized wave is susceptible to the relativistic Raman and the modulational instabilities \cite{Guerin95}, the analysis of these solitary structures is fully justified. First, the present work extends the discussion about solitary waves with NVBC, that until now was limited to the narrow velocity range, $0<V/c<0.051$  \cite{Farina02,Farina05}. Second, the solitary waves could play a role in processes that are faster than the inverse of the parametric instability growth rate. We also point out that any damping mechanism, not included in the present analysis for simplicity, could reduce or suppress these instabilities.

From the mathematical point of view, the problem of finding solitons reduces, through the traveling wave ansatz, to that of finding orbits of an associated Hamiltonian system for $a$ and $\phi$ consistent with the boundary conditions of the fluid model for each type of soliton. The Hamiltonian system is four dimensional, time-reversible and autonomous. It admits four fixed points that, for brevity, will be denoted  by the letter $Q$ and the value of $\phi$ and $a$ inside brackets. Only three of them, $Q_0^\pm=(0,\pm a_0)$ and $Q_1=(\phi_0, 0)$,  play a role when discussing the existence of solitary waves.  Orbits that are asymptotic to an equilibrium 
as $x\rightarrow \pm\infty$ are called \emph{homoclinic connections}, while orbits that connect two distinct equilibria
in the limits $x\rightarrow \pm\infty$ are termed \emph{heteroclinic connections} (see \cite{Champneys98} for a review of homoclinic orbits in reversible systems).  Therefore, grey solitons, dark solitons and shock waves correspond to the connecting orbits $Q_0^{\pm}\hetArr Q_0^{\pm}$,  $Q_0^+\hetArr Q_0^-$ and 
$Q_0^{\pm}\hetArr Q_1$ respectively. 

Existence and robustness under variations in $V$ and $\omega$ of such connecting orbits can be directly inferred by simple geometrical arguments from the theory of dynamical systems, with a critical role played by the properties of the equilibria of the system and in particular their linear stability.
Previous works \cite{Farina02,Farina05} studied the parametric domain where $Q_0^{\pm}$ is a saddle-center, 
with eigenvalues $\lambda_{1,2}=\pm\alpha$, $\lambda_{3,4}=\pm i\beta$, and they found solutions in the velocity range $0<V/c<0.051$.

Here, we extend the analysis  to velocities up to $V/c=1$, and specifically in regions of the parametric domain where $Q_0^{\pm}$ is a saddle-focus, \ie\ its eigenvalues are of the form $\pm(\alpha\pm i\beta$). It is well known \cite{Devaney76} that homoclinic connections of a saddle-focus in Hamiltonian systems do generally exist and are robust against small perturbations of the Hamiltonian. Therefore one expects them to appear in continuous families in the $V-\omega$ plane. Moreover, existence of such a homoclinic orbit, for instance one that corresponds to a  one-hump soliton, implies existence of infinitely many of them, corresponding to solitons with a different number of humps \cite{Devaney76}. Taking into account these insights from the theory of dynamical systems, we numerically identify new families of single and multi-hump solitons of the grey and dark varieties that can be interpreted as counterparts to the branches reported for VBC \cite{Farina01a}. We also report, for the first time, asymmetric solitons in this system and explain the disappearance of certain families of solutions as parameters are varied.

The organization of the paper is as follows. In section \ref{Sec:Dynamical} the system of equations that governs the dynamics of the solitons is revisited, while its fixed points and their possible connecting orbits are discussed in section \ref{Sec:FP:Orbits}. Section \ref{Sec:Algorithm} describes the spectral algorithm \cite{Liu94} that we have used to find homoclinic and heteroclinic orbits, as well as some optimizations carried out for our specific problem. The numerically computed families of dark and grey, symmetric and asymmetric solitons are presented in section \ref{Sec:Results}. In section \ref{Sec:Stability} the stability of certain types of grey solitons is investigated with the aid of particle-in-cell (PIC) simulations. The conclusions are summarize in  section \ref{Sec:Conclusions} where the similarities and differences with the VBC waves are stressed.

\section{\label{Sec:Dynamical}Dynamical equations}

This section briefly summarizes the theory of one-dimensional circularly polarized solitons (see  \cite{Kozlov79,Farina01a,Farina01b} for a detailed discussion). The plasma is assumed to be cold and composed of electrons and ions,  denoted by the subscript $\alpha=e,i$ respectively. It is convenient to use length, time, velocity, momentum, vector and scalar potential, and density  normalized over $c/\omega_{pe}$, $\omega_{pe}^{-1}$, $c$, $m_\alpha c$, $m_ec^2/e$ and $n_0$ respectively. Here $n_0$, $m_\alpha$ and  $\omega_{pe}=(4\pi n_0e^2/m_e)^{1/2}$  are the unperturbed density, the rest mass and the electron plasma frequency. Using this notation the Maxwell (in the Coulomb gauge) and plasma equations  read

\numparts
\begin{equation}
\Delta\mathbf{A}-\frac{\partial^2 \mathbf{A}}{\partial t^2}-\frac{\partial }{\partial t}\nabla\phi=n_e\mathbf{v}_e-n_i\mathbf{v}_i\label{Eq:A:vec}
\end{equation}
\begin{equation}
\Delta\phi=n_e-n_i\label{Eq:Poisson}
\end{equation}
\begin{equation}
\frac{\partial n_\alpha}{\partial t}+\nabla\cdot\left(n_\alpha\mathbf{v}_\alpha\right)=0\label{Eq:continuity}
\end{equation}
\begin{equation}
\frac{\partial \mathbf{P}_\alpha}{\partial t}-\mathbf{v}_\alpha\times\left(\nabla\times\mathbf{P}_\alpha\right)=-\nabla\left(\epsilon_\alpha\phi+\gamma_\alpha\right)\label{Eq:momentum}
\end{equation}\label{Sys:General}
\endnumparts
\noindent where $\mathbf{A}$ and $\phi$ are the vector and scalar potential, $\mathbf{P}_\alpha\equiv\mathbf{p}_\alpha+\epsilon_\alpha\mathbf{A}$, $\gamma_\alpha\equiv(1+|\mathbf{p}_\alpha|^2)^{1/2}$ and $\mathbf{p}_\alpha$ and $\mathbf{v}_\alpha\equiv\mathbf{p}_\alpha/\gamma_\alpha$ are the kinetic momentum and the fluid velocity respectively. For convenience the dimensionless parameter $\epsilon_\alpha \equiv (q_\alpha m_e)/(e m_\alpha) $ has been introduced ($q_e=-e$ and $q_i=e$ are species charges).

Taking $\displaystyle{\partial_y=\partial_z=0}$, the Coulomb Gauge immediately gives $\displaystyle{\mathbf{A}=\mathbf{A}_\perp}$ where  $\displaystyle{\perp}$ denotes the direction perpendicular to  $\displaystyle{x}$. The transverse component of  \eref{Eq:momentum} yields $\displaystyle{\mathbf{P}_{\perp\alpha}=0}$. We also take all the variables to  be functions of $\displaystyle{\xi=(x-Vt)/\sqrt{1-V^2}}$ and assume a circularly polarized vector potential
\begin{equation}
A_y+iA_z= a(\xi)e^{\frac{i \left(kx-\omega t\right)}{\sqrt{1-V^2}}}\label{EQ:Ayz}
\end{equation}
Here, $V$ is the group velocity of the solitary wave, $k$ is the wavevector and $\omega$ is the frequency in a frame moving with the solitary wave. With the above assumptions, equation  \eref{Eq:continuity}  and the longitudinal component of \eref{Eq:momentum} can be integrated. Imposing the boundary conditions $a=\pm a_0$, $\phi=0$, $n_\alpha=1$ and $p_{x\alpha}=0$ as $x\rightarrow -\infty$, the kinetic momentum, energy and the density of each species are functions of just the potentials $\phi$ and $a$. For instance, the densities and the $\gamma$ factors are given by
\begin{equation}
n_\alpha(\phi,a)=\frac{V\left(\psi_\alpha-Vr_\alpha\right)}{(1-V^2)r_\alpha}\label{Eq:density}
\end{equation}
\begin{equation}
\gamma_\alpha(\phi,a) = \frac{\psi_\alpha-Vr_\alpha}{1-V^2}\label{Eq:gamma}
\end{equation}
with $\psi_\alpha\equiv \Gamma_{\alpha}-\epsilon_\alpha\phi$, $r_\alpha\equiv\left[\psi_\alpha^2-(1-V^2)(1+\epsilon_\alpha^2a^2)\right]^{1/2}$ and $\Gamma_{\alpha }=(1+\epsilon_\alpha^2a_0^2)^{1/2}$. For brevity we write $\epsilon_i\rightarrow \epsilon$ and for the numerical calculation we set  $\epsilon=1/1836$. Substituting in  \eref{Eq:A:vec} and \eref{Eq:Poisson} yields  \cite{Farina02,Farina05}

\numparts
\begin{equation}
a''=\left[V\left(\frac{1}{r_e}+\frac{\epsilon}{r_i}\right)-\omega^2\right]a\label{sys:Hamiltonian:a}
\end{equation}
\begin{equation}
\phi''=V\left(\frac{\psi_e}{r_e}-\frac{\psi_i}{r_i}\right)\label{Eq:phi}\label{sys:Hamiltonian:b}
\end{equation}
\endnumparts
and $V=\omega/k$, where the prime denotes derivative with respect to $\xi$. We remark that, for solutions described by equation \eref{EQ:Ayz}, the group velocity of the solitary wave $V$ should be equal to the phase velocity $\omega/k$. Solutions with different velocities require the addition of a relative phase in equation  \eref{EQ:Ayz} \cite{Farina02,Farina05}. 

System \eref{sys:Hamiltonian:a}-\eref{sys:Hamiltonian:b} describes the dynamics of  localized electromagnetic modulations trapped by a self-generated Langmuir wave embedded in a infinitely long electromagnetic wave. Introducing the momenta $P_a=(1-V^2)a'$ and $P_{\phi}=-\phi'$, it can be written as a fourth order Hamiltonian system with Hamiltonian  
\begin{equation}
\fl
H(a,P_a,\phi,P_\phi)=\frac{1-V^2}{2}\left[\left(\frac{P_a}{1-V^2}\right)^2+\omega^2a^2\right]-\frac{1}{2}P_\phi^2
+V\left[r_e(a,\phi)+\frac{r_i(a,\phi)}{\epsilon}\right]\label{Eq:Inv}
\end{equation}
We also note that the conditions $r_\alpha^2>0$ imply the following restriction on $\phi$ \cite{Bonatto07}
\small                       
\begin{equation}
\sqrt{(1-V^2)(1+a^2)}-\Gamma_{e}<\phi<\frac{1}{\epsilon}\left(\Gamma_{i}-\sqrt{(1-V^2)(1+\epsilon^2a^2)}\right)\label{Eq:bound:general}
\end{equation}
\normalsize  

\section{\label{Sec:FP:Orbits}Fixed points and connecting orbits}

\subsection{\label{Subsection:FP:stability} Fixed points and their stability}

The fixed points  are obtained by setting the righthand side of  \eref{sys:Hamiltonian:a}- \eref{sys:Hamiltonian:b} equal to zero and, for brevity, they will be denoted by  $Q(\phi,a)$ (assuming  $\phi'=a'=0$ as $\xi\rightarrow \pm\infty$). We note that only orbits connecting the fixed points $Q_0^{\pm}=(0,\pm a_0)$ are consistent with the previously imposed boundary conditions. Hence, $Q_0^{\pm}$ and their stability properties play an important role when discussing the presence of solitary waves in the full fluid system. 

\subsubsection{Fixed points  $Q_0^\pm$.}

One readily verifies that $Q_0^\pm=(0,\pm a_0)$ is a fixed point of system   \eref{sys:Hamiltonian:a}-\eref{sys:Hamiltonian:b}  if the following dispersion relation is satisfied
\begin{equation}
\omega^2=\frac{1}{\Gamma_{e}}+\frac{\epsilon}{\Gamma_{i}}\label{Eq:a0}
\end{equation}
Its physical interpretation is evident when we express  \eref{Eq:a0} as a function of the wavevector $k_{LF}$ and the frequency in the laboratory frame $\omega_{LF}$ \cite{Farina01a,Farina05}
\begin{equation}
\omega^2_{LF}=k_{LF}^2+\frac{1}{\Gamma_{e}}+\frac{\epsilon}{\Gamma_{i}}
\end{equation}   
that is the normalized dispersion relation of a pure transverse electromagnetic wave with relativistic effects and the ion motion. Note that equation \eref{Eq:a0} has a solution within the range $0<\omega^2<1+\epsilon$.

 The stability of  $Q_0^{\pm}$ is determined by the eigenvalues of the Jacobian of system   \eref{sys:Hamiltonian:a}-\eref{sys:Hamiltonian:b}. In the $V-\omega$  plane there exist 3 different regions of stability; for any given $\omega$, $Q_0^{\pm}$ is a saddle-center if $V<V_{SC}$, a center if $V_{SC}<V<V_{SF}$ and a saddle-focus if $V>V_{SF}$. The velocities $V_{SC}$ and $V_{SF}$  are (see Appendix \ref{Sec:Appendix} and figure \ref{Fig:Parametric})
\small
\begin{eqnarray}
V_{SC} &\equiv& \sqrt{1+\frac{\left[2a_0(1-\epsilon^2)\Gamma_e\Gamma_i\right]^2}{(\Gamma_i^3+\epsilon\Gamma_e^3)^2-\left[\Gamma_i^3(\Gamma_e^2+a_0^2)+\epsilon\Gamma_e^3(\Gamma_i^2+\epsilon^2a_0^2)\right]^2}}\nonumber\\
&&\label{Eq:Vsc}\\
V_{SF} &\equiv& \sqrt{1-\left[\frac{2a_0(1-\epsilon^2)\Gamma_e\Gamma_i}{\Gamma_i^3(\Gamma_e^2+a_0^2)+\epsilon\Gamma_e^3(\Gamma_i^2+\epsilon^2a_0^2)}\right]^2}\label{Eq:Vsf}
\end{eqnarray}
\normalsize
  
Note that $V_{SF}\sim V_{SC}\rightarrow 1$ as $a_0\rightarrow \infty$ ($\omega\rightarrow 0$) and  $V_{SF}\rightarrow 1$ and $V_{SC}\rightarrow \epsilon/(1-\epsilon+\epsilon^2)$ as $a_0\rightarrow 0$ ($\omega^2\rightarrow 1+\epsilon$).

\normalsize
\subsubsection{Fixed point  $Q_1$.}

\begin{figure}[h]
\begin{center}
\includegraphics[scale=0.43,clip=true]{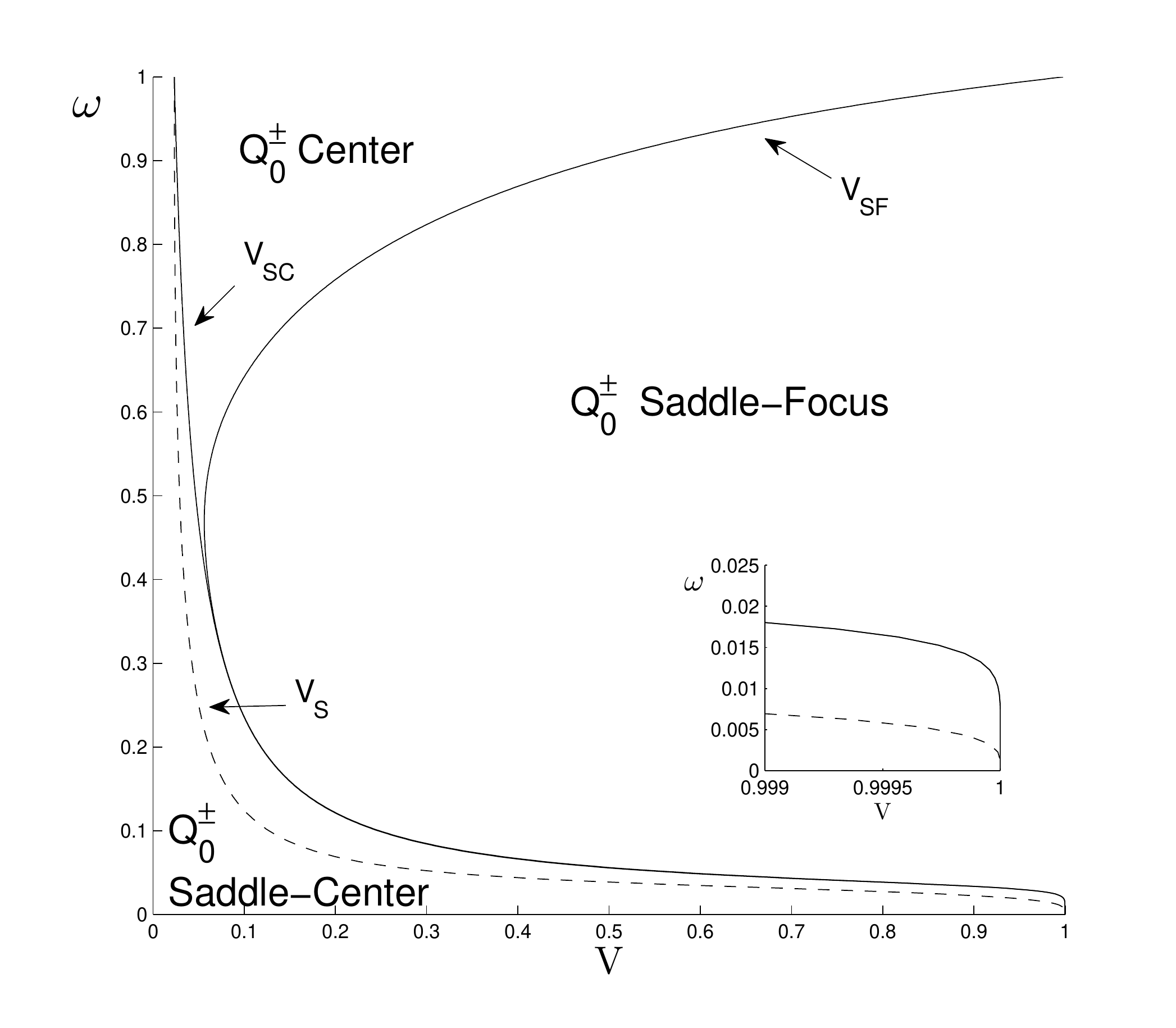}
 \caption{\label{Fig:Parametric}
		Characteristic velocity curves and $Q_0^{\pm}$ stability regions in the $V-\omega$ plane.
                The velocities $V_{SC}(\omega)$ and $V_{SF}(\omega)$ (solid lines) split the plane in three 
		regions where $Q_0^{\pm}$ has different stability. Along the curve $V_S(\omega)$ (dashed line)
		shock waves are possible. The inset shows a detail close to $V\sim1$ where both $V_{SF}$ 
		and $V_S$ approach 1 as $\omega\rightarrow 0$.
}
\end{center}
\end{figure}

System  \eref{sys:Hamiltonian:a}-\eref{sys:Hamiltonian:b}  also admits the fixed point $Q_1=((\Gamma_{0i}-\Gamma_{0e})/(1+\epsilon),0)$. Since orbits must connect $Q_1$ to $Q_0^{\pm}$, we need to enforce the same value of the invariant (equation  \eref{Eq:Inv}) at both fixed points. This yields 
\begin{equation}
V_S = \left[\left(1-\frac{\epsilon a_0^2}{2\Gamma_{e}\Gamma_{i}}\right)^2-\left(\frac{1+\epsilon}{\Gamma_{i}+\epsilon\Gamma_{e}}\right)^2\right]^{-1/2}\frac{\epsilon a_0^2}{2\Gamma_{e}\Gamma_{i}}\label{Eq:Vs}
\end{equation}
which gives an explicit relation between the velocity of the wave and the asymptotic value $a_0$, plotted in figure \ref{Fig:Parametric}. As $a_0\rightarrow 0$ $(\omega^2\rightarrow 1+\epsilon)$ one has $V_S\rightarrow \epsilon/(1-\epsilon+\epsilon^2)$. Appendix  \ref{Sec:Appendix} shows that $Q_1$ is a saddle-center for velocities given by  \eref{Eq:Vs} and, in principle, shock waves are possible in the range $\epsilon/(1-\epsilon+\epsilon^2)<V<1$. Shock waves belong to the parametric domain where $Q_0^{\pm}$ is a saddle-center, see figure \ref{Fig:Parametric}, and will not be studied in the present work (see   \cite{Farina02} and \cite{Farina05}). 

\subsubsection{Fixed point $Q_2$.}

The system has an additional fixed point $Q_2$ for velocities $V$ greater than
\begin{equation} 
v\equiv\frac{\Gamma_{i}+\epsilon\Gamma_e}{1+\epsilon}\sqrt{\frac{(\Gamma_{i}+\epsilon\Gamma_e)^2-(1+\epsilon)^2}{(1+\epsilon)^2\Gamma_e^2\Gamma_i^2-(\Gamma_{i}+\epsilon\Gamma_e)^2}}\label{Eq:V:Q2}
\end{equation}
(see appendix \ref{Sec:Appendix}). As $Q_0^{\pm}$ and $Q_2$ must share the same value of $H$ for a connection to exist, a condition similar to  \eref{Eq:Vs} can be obtained, which cannot be fullfilled for any value of the parameters and therefore heteroclinic connections involving $Q_2$ and $Q_0^{\pm}$ are not possible (see Appendix \ref{Sec:Appendix}).

\subsection{\label{Subsection:Connecting} Connecting orbits}

We now turn to the discussion of the conditions under which heteroclinic and homoclinic connections, and therefore solitons of the fluid system \eref{Eq:A:vec}-\eref{Eq:momentum} are expected to exist. We will need to recall well known facts from the theory of dynamical systems (see \cite{Champneys98}, for example) 
and introduce the notion of the \emph{stable (unstable) manifold} $W^s_{i}$ ($W^u_{i}$) of a fixed point $Q_i$, as the set of forward (backward) in $\xi$ trajectories that terminate at $Q_i$. A heteroclinic (homoclinic) connection from $Q^i$ to $Q^j$ (to itself) lies on the intersection of the unstable manifold of $Q^i$ and the stable manifold of $Q^j$ ($Q^i$).

Stable and unstable manifolds are complicated objects with intriguing structure and their visualization can become a formidable task (see \cite{krauskopf05} for a review of methods). Here it will be enough to consider the dimensionality of stable and unstable manifolds, as this dictates whether or not they are in general expected to intersect. We note that dynamics of the $4$-dimensional system  \eref{sys:Hamiltonian:a}-\eref{sys:Hamiltonian:b}  are constrained to a $3$-dimensional \emph{energy-manifold} by conservation of the invariant $H$, equation \eref{Eq:Inv}. The intersection of stable and unstable manifolds has to take place within this energy-manifold. 

We first discuss homoclinic connections involving $Q_0^{\pm}$. When $Q_0^{\pm}$ is a saddle-center, both
its stable and unstable manifolds are $1$-dimensional and are, in general, not expected to intersect
in $3$-dimensional space. If they do intersect, they necessarily have to be the same from uniqueness of solutions. A homoclinic connection is therefore not generic, in the sense that it requires the condition $W^s_0=W^u_0$ which can only be fullfilled for specific $V$ and $\omega$ values. Hence the homoclinic connections in the saddle-center case are expected to occur in branches in the $V-\omega$ plane. However, the existence of a continuous spectrum of single-hump solitons when $Q_0^{\pm}$ is a saddle-center has been suggested for VBC \cite{Poornakala02} and NVBC \cite{Farina02,Farina05} boundary conditions. Unless further restrictions are imposed, such a continuous spectrum has to be interpreted as a numerical artifact, only valid within the accuracy of long-time integration of system  \eref{sys:Hamiltonian:a}-\eref{sys:Hamiltonian:b}. Nevertheless, singular cases could arise. For instance, at the $V=0$ case where the dynamics is $2$-dimensional, it can be proved analytically that standing solitary wave solutions exist within a continuous $\omega$ range \cite{Esirkepov98}.

In the saddle-focus case on the other hand, $W^s_0$ and $W^u_0$ are $2$-dimensional and are in general expected to intersect transversally along a $1$-dimensional curve in the $3$-dimensional energy-manifold. Therefore we expect homoclinic solutions to exist generically in the $V-\omega$ plane. Furthermore it is well known that, for given parameters, if one such transverse intersection exists then there exist infinitely many, which have been shown  to form a local, complete Horseshoe structure \cite{Devaney76}. In our case this implies that for any one-hump soliton we can find an associated family of multi-hump solitons for any given $V$ and $\omega$ \cite{Champneys99}.

The symmetry properties of  \eref{sys:Hamiltonian:a}-\eref{sys:Hamiltonian:b}  help us deduce more properties of the 
connecting orbits. Equations  \eref{sys:Hamiltonian:a}-\eref{sys:Hamiltonian:b}  are invariant under the reflection symmetry, 
$a\rightarrow -a,\, a'\rightarrow -a'$. As a result, for any homoclinic 
$Q_0^{+}\hetArr Q_0^{+}$ connection there exists an identically shaped 
$Q_0^{-}\hetArr Q_0^{-}$ one. Note that, since we impose inhomogeneous boundary
conditions, we cannot have reflection invariant connections.
Moreover, reflection symmetry allows us to carry
many of the results available for homoclinic orbits over to heteroclinic orbits
involving  $Q_0^+$ and $Q_0^-$ (dark solitons); the two equilibria 
can be considered as a single equilibrium in a reduced system in which $a$
is identified with $-a$ and in which heteroclinic connections become
homoclinic \cite{Champneys99}. Therefore we can expect heteroclinic orbits in the saddle-focus parametric regime to occur
generically. Furthermore, the $Q_0^+\hetArr Q_0^-$ connection has an identically shaped 
$Q_0^-\hetArr Q_0^+$ counterpart. 

System  \eref{sys:Hamiltonian:a}-\eref{sys:Hamiltonian:b}  is also time-reversible, that is invariant
under simultaneous change of sign of $\xi$ and the generalized momenta $P_\phi$, $P_a$. As a result,
homoclinic orbits can either be symmetric (self-dual, $\phi(\xi)=\phi(-\xi),\, a(\xi)=a(-\xi)$) 
or come in pairs of asymmetric orbits related by time-reversal. Similarly, heteroclinic connections 
$Q_0^+\hetArr Q_0^-$ have to be either antisymmetric functions of $\xi$ ($\phi(\xi)=\phi(-\xi),\, a(\xi)=-a(-\xi)$) 
or come in asymmetric pairs, related by $\xi$-reversal. Note that the antisymmetric heteroclinic solitons
are associated with the combined action of $\xi$-reversal and reflection. Asymmetric homoclinic and heteroclinic orbits
in $\xi$-reversible Hamiltonian systems are well studied \cite{Champneys98}, but to our knowledge 
they appear here for the first time in the context of relativistic solitons, see section \ref{Sec:Results}.
The addition of a perturbation that breaks the conserved quantity $H$ but preserves 
the reversibility would destroy asymmetric solutions \cite{Champneys98}.

For later use, we introduce the so-called \emph{symmetric section} $\mathcal{S}:\,a'=\phi'=0$, which is in a sense a symmetry hyperplane of the system, as any point in it is left invariant under $\xi$-reversal.  
Symmetric homoclinic orbits have to intersect $\mathcal{S}$ \cite{Champneys98}. 
As we will see in section \ref{Sec:Results} symmetric homoclinic orbits can disappear through a mechanism 
referred  to as \emph{coalescence} \cite{Champneys98}, as a result of which the stable and unstable manifolds fail to intersect on the symmetric section $\mathcal{S}$.

\section{Numerical algorithm}\label{Sec:Algorithm}

Due to time reversibility, the computation of symmetric homoclinic or antisymmetric heteroclinic orbits of  \eref{sys:Hamiltonian:a}-\eref{sys:Hamiltonian:b}   involves the solution of a boundary value problem on the semi-infinite interval ($-\infty,0$). Within the regime where $Q_0^{\pm}$ is a saddle-center and the unstable manifold is one-dimensional, previous works truncate this interval and consider the linearized dynamics close to the fixed points, solving for the parameter values in the $V-\omega$ plane in which the boundary value problem is satisfied \cite{Kozlov79,Farina01a,Farina02,Farina05}. For the saddle-focus case, where the local unstable manifold is a $2$-dimensional plane, one would also need to solve for the initial condition in this plane, for instance by parametrizing it by a polar angle \cite{Champneys93}. 
 
Here we implement the rational spectral collocation algorithm of Ref. \cite{Liu94} that avoids both the truncation of the domain and the introduction of an additional parameter. Let us consider the system  $\mathbf{u}''=\mathbf{f}(\mathbf{u})$ with  $\mathbf{u}\equiv(a,\ \phi)^T$ and the components of $\mathbf{f}$ given by the right-hand sides of   \eref{sys:Hamiltonian:a}- \eref{sys:Hamiltonian:b}. Assuming a fast enough approach of the solutions  to the asymptotic values  $\mathbf{u}\rightarrow \mathbf{u}_{\pm}$  as $\xi\rightarrow \pm\infty$  \cite{Shen09},  the variables can be expanded as a sum of orthogonal rational functions
\begin{eqnarray}
u_i(\xi) = \sum_{k=0}^{M+1}C_{ik}\cos\left[k\cot^{-1}(\xi)\right], \ \ \ \ \ \ i=1,\ 2  \label{Eq:expansion1}
\end{eqnarray}

The basis functions suggest the following choice for the M collocation points
\begin{eqnarray}
\xi_j = \cot\left(\frac{j\pi}{M+1}\right) && 1\le j\le M
\end{eqnarray}
that are complemented by the two following collocation points $\xi_0=+\infty$ and $\xi_{M+1}=-\infty$. 
The coefficients $C_{ik}$ are given by
\begin{equation}
C_{ik}=\frac{2}{(M+1)\bar{c}_k}\sum_{m=0}^{M+1}\frac{u_{i}(\xi_m)}{\bar{c}_m}\cos\left(\frac{mk\pi}{M+1}\right) \ \  \small 0\le k\le M+1
\end{equation}
with $\bar{c}_m=2$ if $m=0$ or  $m=M+1$ and $\bar{c}_m=1$ if $1\le m\le M$.

Computing the second derivative from   \eref{Eq:expansion1} and substituting the results  in $\mathbf{u}''=f(\mathbf{u})$  yield $2M$ nonlinear algebraic equations
\begin{equation}
\sum_{k=0}^{M+1}D_{jk}C_{ik}+f_{i}[\mathbf{u}(\xi_j)]=0\ \ \ \ \ \ 1\leq j\leq M,\ \ \ \ \ \ i=1,2\label{Eq:collocation:eqs}
\end{equation}
where 
\begin{equation}
\small D_{jk}\equiv k\left[k\cos\left(\frac{kj\pi}{M+1}\right)+\frac{2\sin\left(\frac{kj\pi}{M+1}\right)}{\tan\left(\frac{j\pi}{M+1}\right)}\right]\sin^4\left(\frac{j\pi}{M+1}\right)
\end{equation}
We also have 4 boundary conditions 
\numparts
\begin{equation}
 \sum_{k=0}^{M+1}C_{ik}= u_i^+\label{Eq:collocation:bc1}
\end{equation}
\begin{equation}
 \sum_{k=0}^{M+1}(-1)^{k}C_{ik}= u_i^-\label{Eq:collocation:bc2}
\end{equation}

\endnumparts
with  $1\le i\le 2$.

Since the system is autonomous, every translation on $\xi$ of its solution is also a solution. This indeterminacy can be removed by adding the phase condition
\begin{equation}
\int_{-\infty}^{+\infty}<\mathbf{u}'(\xi)-\tilde{\mathbf{u}}'(\xi),\mathbf{u}''(\xi)>=0\label{Eq:collocation:phase}
\end{equation}
where $<\cdot,\cdot>$ denotes the $l^2$ inner product and $\tilde{\mathbf{u}}$ the previous orbit on a branch. Solutions occuring at branches $\omega=\omega(V)$ need this extra condition to adequately vary the parameter. This would be the case of the branches found for $V<V_{SC}$ \cite{Farina02,Farina05}. However, as we will see, for  $V>V_{SF}$ the solutions exist continuously in the $V-\omega$ plane. In this case  equations \eref{Eq:collocation:eqs} and \eref{Eq:collocation:bc1}-\eref{Eq:collocation:bc2} constitute a set of $2(M+2)$ equations for the $2(M+2)$ coefficients $C_{ik}$ that can be solved for fixed values of $\omega$ and $V$.  We used a Newton-Raphson algorithm with  its Jacobian calculated analytically to speed up the convergence.

Since  system \eref{sys:Hamiltonian:a}-\eref{sys:Hamiltonian:b}  does  not include the first derivatives of the variables $(\phi,a)$ on the right-hand side, we have directly computed the second derivative. This is a difference from the general scheme given in reference \cite{Liu94} that allows us to divide by two the number of coefficients. Note also that for asymmetric solutions all the coefficients must be calculated whereas in the case of symmetric or antisymmetric solutions just one half of the coefficients are needed (for instance symmetric grey solitons are even solutions and $C_{i,2k+1}=0$). 

The variable $\xi$ can be stretched according to $\xi\rightarrow L\xi$. Here $L$ is a scaling factor that can be used to optimize accuracy  of the solution. Although some strategies have been discussed to  find a proper value of $L$ \cite{Boyd82}, in practice it is chosen by experimentation with different $L$ for a given value of $M$ (as suggested in reference \cite{Boyd01}). We found $L=6$ to be an adequate value for our numerical computations and a number of collocation points $M\sim 300$. The initial guess for the Newton-Raphson algorithm can be obtained from the intersection of
stable and unstable manifolds, as discussed in section \ref{Sec:Results}. For asymmetric solutions it can also be built up from pieces of symmetric solutions of different parity, smoothed by dropping the high order coefficients. Similarly, guesses for multi-hump solutions can be built up from single hump solutions of appropriate shape. Once we have a good initial guess, the spectral algorithm gives an accurate solution that can be used as initial guess for other parameter values.

We finally remark that the accuracy of the solutions can be checked by testing the spectral convergence: a plot of the logarithm of the coeficients $C_{ik}$ versus the order $k$ should decay linearly. This is an advantage as compared to the truncation of the interval methods where the error is controlled by the initial distance to the fixed point which is fixed arbitrarily. Such a criterion allows to separate true from false solutions, specially when the fixed point has eigenvalues with small real part (in absolute value) and the solutions approach slowly to their asymptotic values. 

\section{Numerical results}\label{Sec:Results}

\subsection{Families of solitary waves}

\begin{figure}[h]
\begin{center}
\includegraphics[scale=0.43]{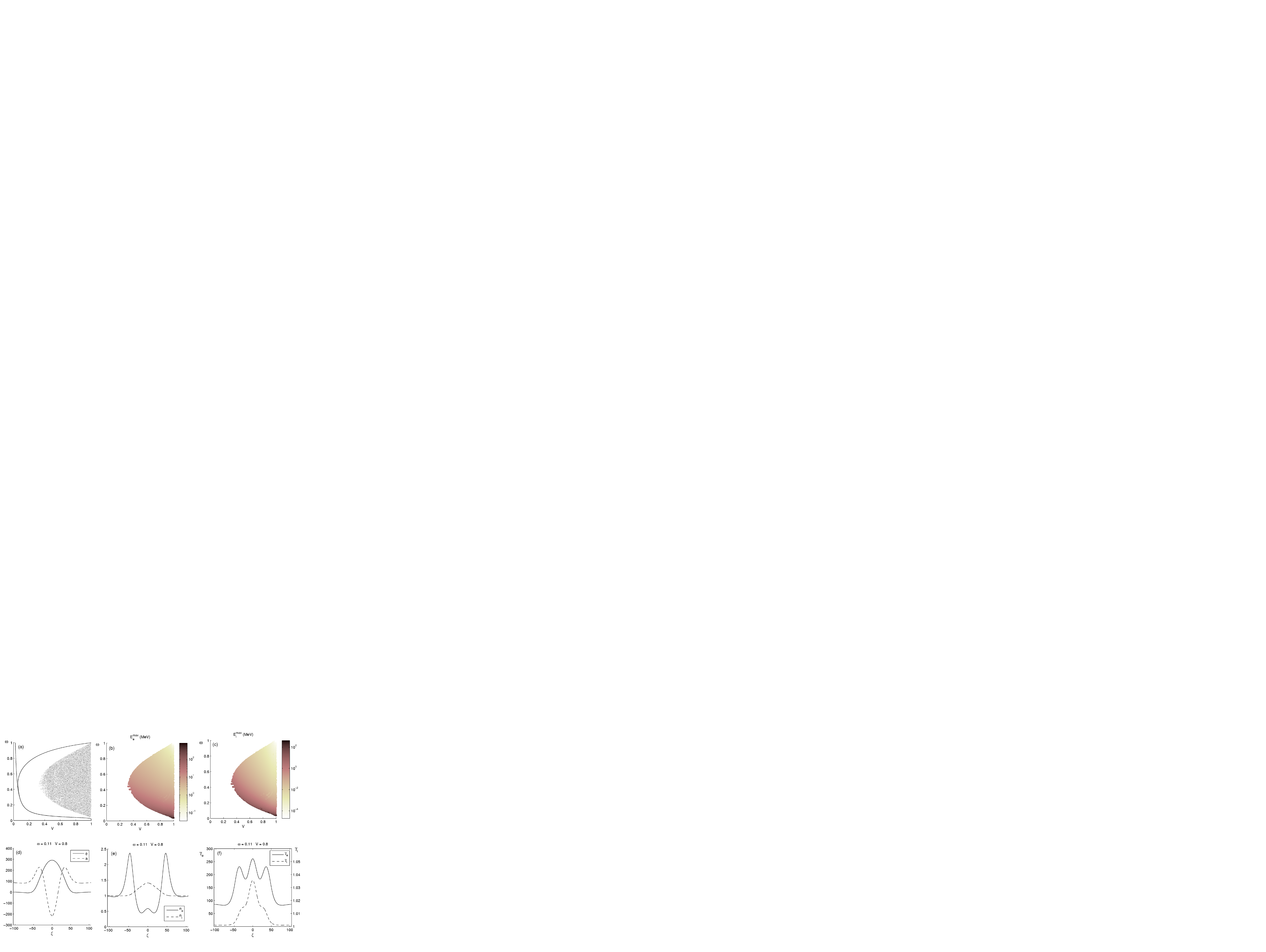}
 \caption{\label{Fig:HOMO1}(Color online) A family of one-hump symmetric grey solitons.}
\end{center}
\end{figure}

\begin{figure}[h]
\begin{center}
\includegraphics[scale=0.43]{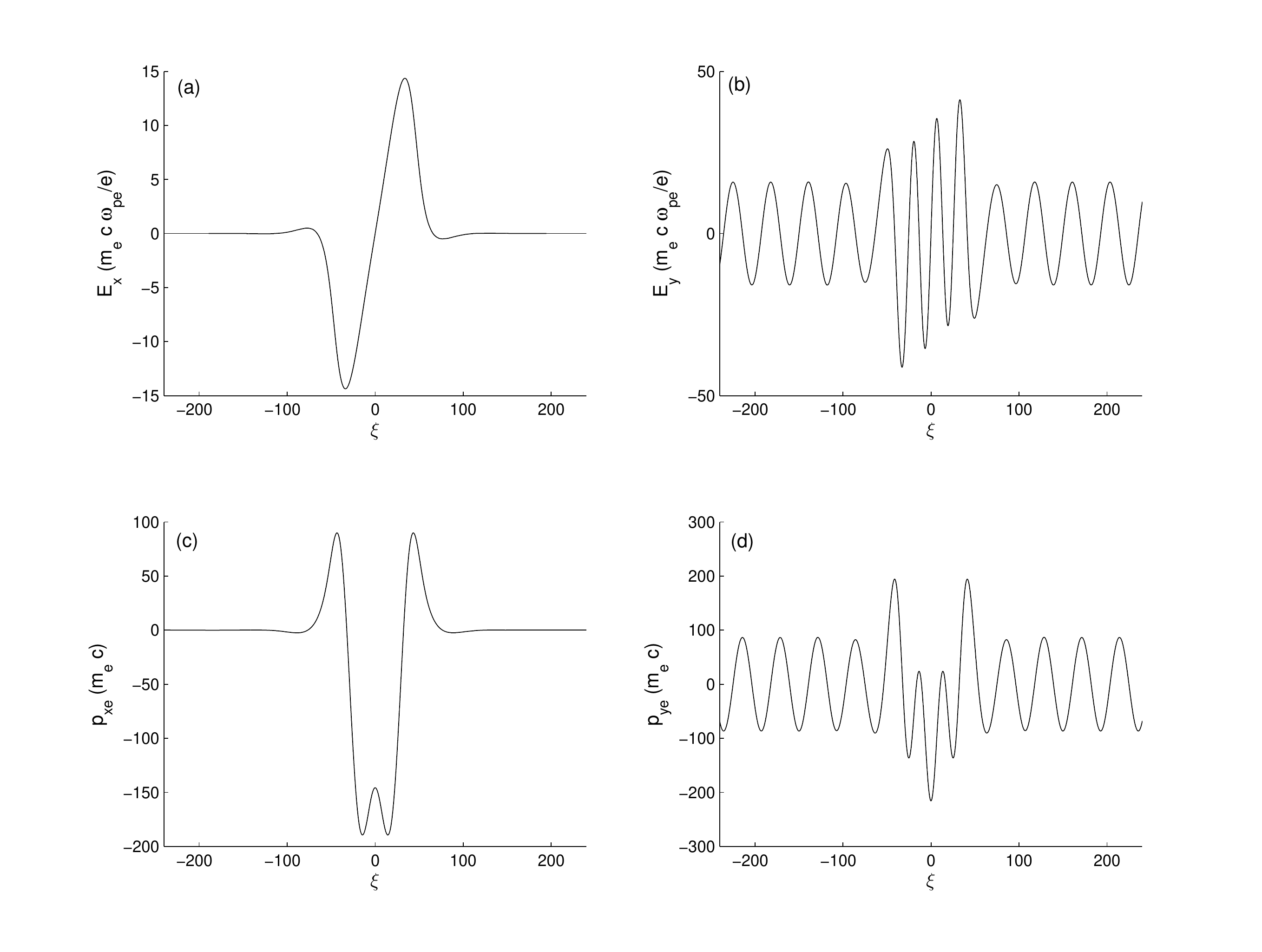}
 \caption{\label{Fig:Grey:Physical} Electric field and electron momenta at a given time for a grey soliton with $V = 0.8$ and $\omega = 0.11$ }
\end{center}
\end{figure}

Figure \ref{Fig:HOMO1} depicts our numerically computed grey soliton solutions, belonging to a certain family. In panel  \ref{Fig:HOMO1}(a)  we also plot the curves $V_{SC}$ and $V_{SF}$. The panels (d)-(f) display a representative solution at the parameter values $V=0.8$ and $\omega = 0.11$, just above  the frequency $\omega\sim 0.10545$ where these solutions disappear for $V=0.8$. The densities and the $\gamma$ factors are computed from equations  \eref{Eq:density} and \eref{Eq:gamma} respectively (note the different scales for $\gamma_e$ and $\gamma_i$). We point out that for this particular family of solutions the ion density and the potential $\phi$ exhibit one hump at the center of the solution. The physical meaning of the solitary waves with NVBC can be undestood with the aid of figure \ref{Fig:Grey:Physical} where we plotted  electric field and electron momenta components. The longitudinal electric field $E_x$ and momentum $p_{xe}$ vanish outside the solitary wave that, as the plots of $E_y$ and $p_{ye}$ show, can be intepreted as a localized modulation of an infinitely long circularly polarized electromagnetic wave ($E_z$ and $p_{ze}$ are not presented).

Even though our fluid description of the plasma cannot take into account discrete particle effects like acceleration or heating, it is interesting to compute  the maximum value of the electron and ion kinetic energy within the solitary wave 
\begin{eqnarray}
E_\alpha^{max} = max[m_\alpha c^2(\gamma_\alpha-1)]
\end{eqnarray}
For the previously discussed family, these quantities are plotted in figure \ref{Fig:HOMO1}(b) and \ref{Fig:HOMO1}(c) (z-axis is in a logarithmic scale). The electron fluid can reach energies of the order of hundreds of MeV and ions 
several tens of MeV. We also note that, for a fixed value of $\omega$, the largest energy is reached at the lowest admissible $V$ value. 

\begin{figure}[h]
\begin{center}
\includegraphics[scale=0.35]{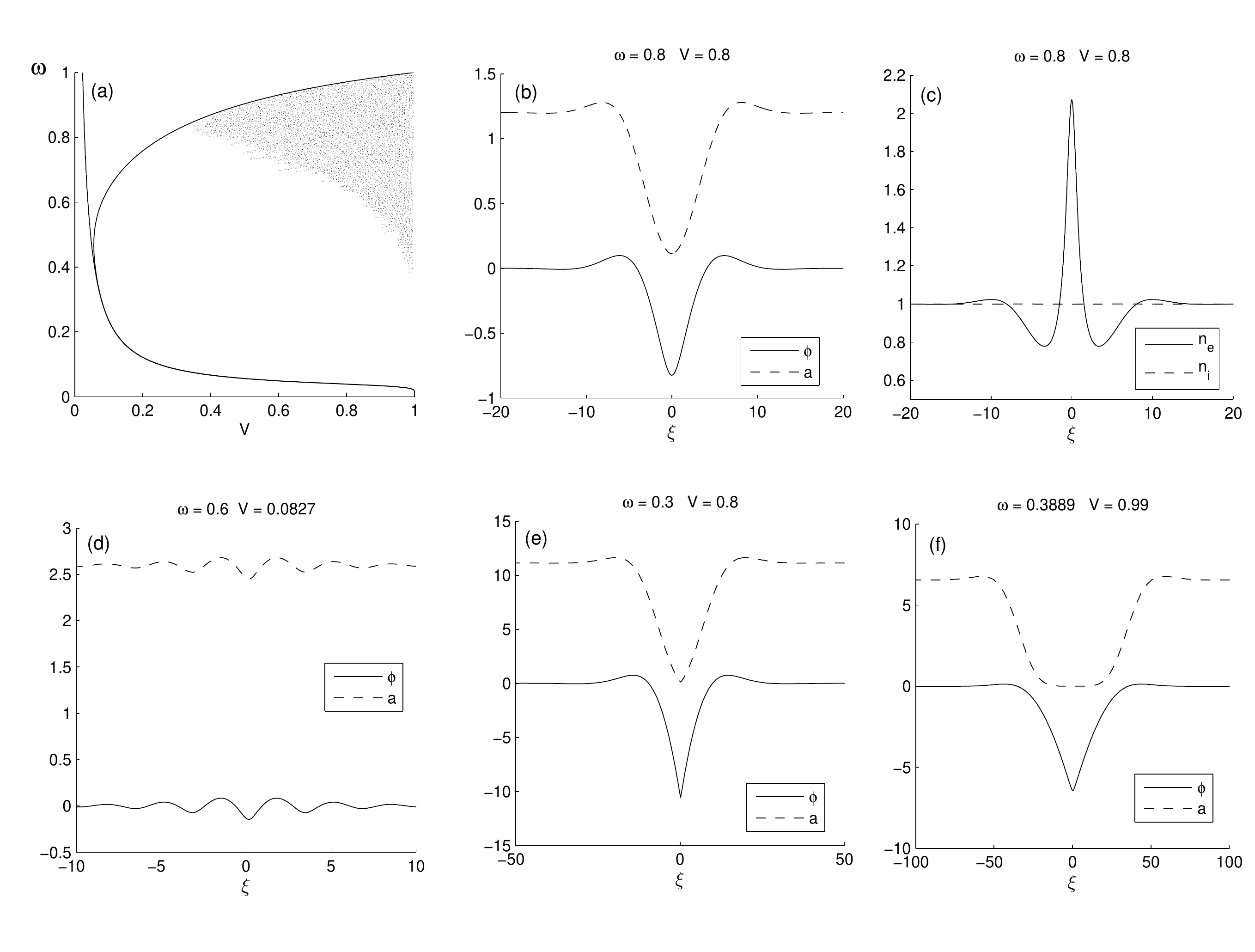}
 \caption{\label{Fig:HOMO2}  A family of one-hump symmetric grey solitons overlapping in the parameter space with the family of figure \ref{Fig:HOMO1}}
\end{center}
\end{figure}

Figure \ref{Fig:HOMO2} corresponds to a different family of grey solitons. In this case both potentials present a minimum at the center of the soliton and the electron density has one central peak. The panel \ref{Fig:HOMO2}(a) shows the existence domain found by randomly varying the parameters $V$ and $\omega$ (as we also did for figure \ref{Fig:HOMO1} ). However, by fixing one of the parameters, using a small step for the other and initializing  the algorithm with the solution obtained at the previous iteration, we were able to find  solitary waves of the same type  for parameter values outside the region indicated in \ref{Fig:HOMO2}(a) (see  figure \ref{Fig:HOMO2}(d)- \ref{Fig:HOMO2}(f)). As the frequency decreases, the potential develops a cusp shape at the center of the wave and the  algorithm requires an initial condition closer to the real solution to ensure  convergence. Hence, the boundary exhibited in figure \ref{Fig:HOMO2}(a)  is a numerical artifact as we will also confirm in section \ref{Sec:Sub:coalescence}. We remark that  potential  profiles with a cusp shape have also been  reported  in solitary waves with VBC for parameter values close to the wavebreaking \cite{Farina01a}.

\begin{figure}[h]
\begin{center}
\includegraphics[scale= 0.40,clip=true]{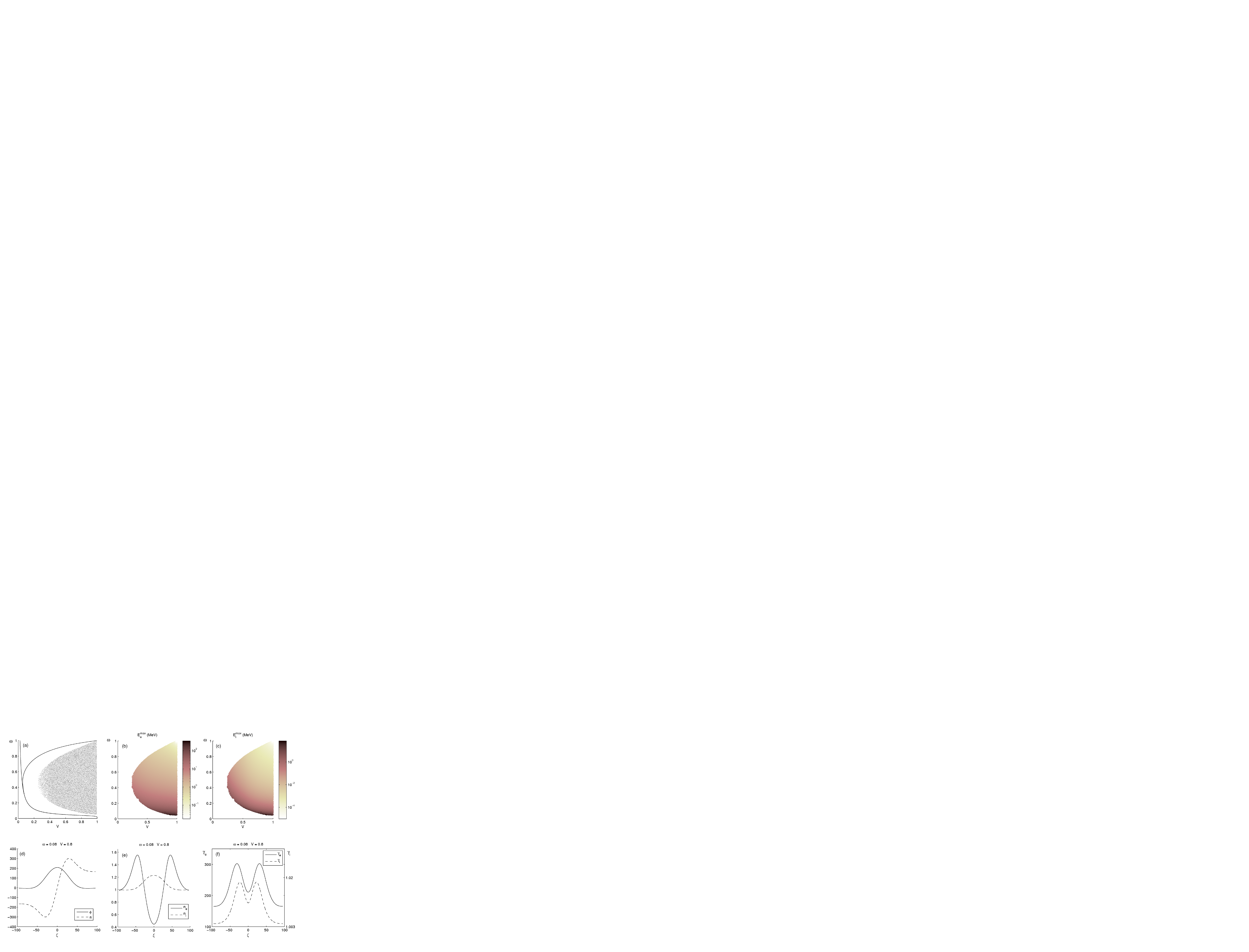}
 \caption{\label{Fig:HETE1}(Color online)  A family of one-hump antisymmetric dark solitons}
\end{center}
\end{figure}

A representative family of dark solitons, or  heteroclinic connections $Q_0^+-Q_0^-$, is shown in figure \ref{Fig:HETE1}. Note that its existence domain encompasses that of the grey solitons of figure \ref{Fig:HOMO1}. Panels \ref{Fig:HETE1}(d)-\ref{Fig:HETE1}(f)  display a particular solution of this family of dark solitons at the parameter values $\omega=0.08$ and $V=0.8$. The potential $\phi$ has one central hump and, as opposed to the grey solitons, the vector potential is an antisymmetric function. The  peak electron and ion kinetic energies are of the order of hundreds and tens of MeV, respectively.

\begin{figure}[h]
\begin{center}
\includegraphics[scale=0.43,clip=true]{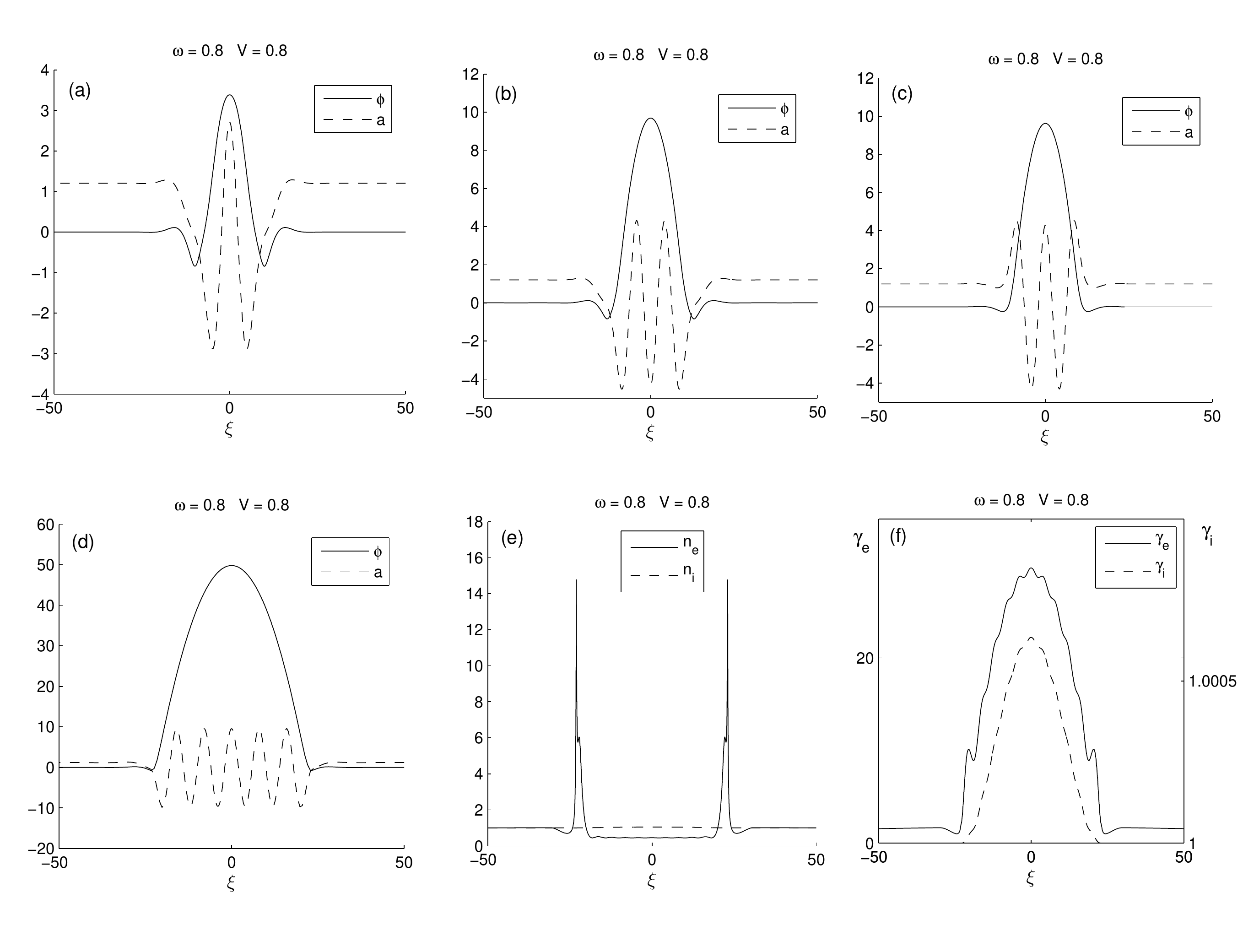}
 \caption{\label{Fig:HOMO3}Some examples of symmetric multi-hump grey solitons.}
\end{center}
\end{figure}

\begin{figure}[h]
\begin{center}
\includegraphics[scale=0.43,clip=true]{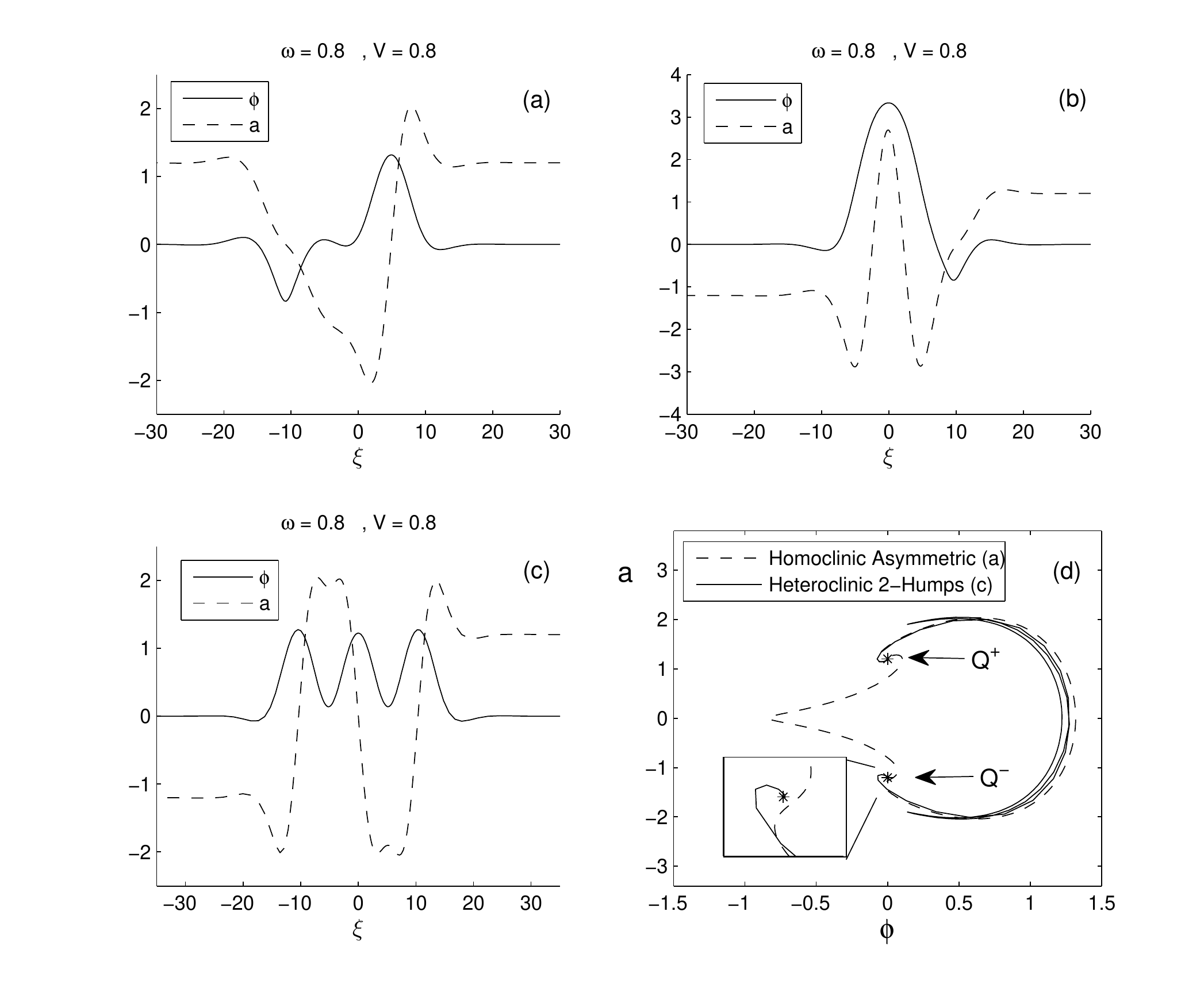}
 \caption{\label{Fig:Multi}Some examples of different types of solutions: (a) homoclinic asymmetric, (b) heteroclinic asymmetric and (c) heteroclinic with 2-humps. Panel (d) shows the projection of orbit (a) and (b) on the $\phi-a$ plane.}
\end{center}
\end{figure}

These three families of solitary waves provide some of the simplest examples of solitary solutions admitted by the system  \eref{sys:Hamiltonian:a}-\eref{sys:Hamiltonian:b}. However, due to the fact that the fixed point is a saddle-focus and the system is Hamiltonian and reversible, multi-hump and asymmetric solutions can exist too. A few examples of grey multi-hump solutions with $\omega= V = 0.8 $ are displayed in figure \ref{Fig:HOMO3}(a)- \ref{Fig:HOMO3}(d). Panels \ref{Fig:HOMO3}(e) and \ref{Fig:HOMO3}(f)  correspond to the density  and $\gamma$ factor  of the solution with 5 humps. The potential $\phi$  exhibits one central hump, while the vector potential has multiple humps. These solutions are characterized by a cavity with a depression of the electron density in which an electromagnetic wave can be trapped.
On the other hand,  grey and dark asymmetric solitons are shown in  figure \ref{Fig:Multi}(a) and \ref{Fig:Multi}(b), respectively. For completeness, Figure \ref{Fig:Multi}(c) also shows a multi-hump dark soliton (heteroclinic connection). Solutions in panels \ref{Fig:Multi}(a) and \ref{Fig:Multi}(c) are plotted in panel \ref{Fig:Multi}(d), where a projection on the phase space $\phi-a$ together with the fixed points $Q^\pm_0$ is shown. Note that the asymmetric grey soliton connects $Q^+_0-Q^+_0$, although it passes close to $Q^-_0$ (see the inset).

\subsection{Coalescence and disappearance of solitary waves in parameter space\label{Sec:Sub:coalescence}}

A striking feature of the families of solitary waves we identified is that they fill in regions with a well defined, within our numerical precision, boundary in the $V-\omega$ plane. As we have seen in section \ref{Sec:FP:Orbits}, homoclinic and heteroclinic orbits lie on the interesection of stable and unstable manifolds of fixed points. This suggests that certain families of orbits can cease to exist when the corresponding manifolds fail to intersect in the neighborhood of the solitary solution. The mechanism responsible for loss of intersection in our case is known as \emph{coalescence} \cite{Buffoni96,Knobloch97,Champneys98} and we now turn to its detailed description, through a numerical experiment involving the disapperance of a member of the family of grey solitons shown in figure \ref{Fig:HOMO1}.

An approximation to the unstable manifold of  $Q_0^+$ can be computed by integrating  equations  \eref{sys:Hamiltonian:a}-\eref{sys:Hamiltonian:b} with initial condition
 \begin{equation} 
[\phi\ \dot{\phi}\ a\ \dot{a}] = [0\ 0\ a_0\ 0]+\sigma\mathbf{v}_1 
\end{equation}
where $\mathbf{v}_1$ is the real part of one of the unstable eigenvectors of $Q_0^+$, associated with an eigenvalue $\tilde{\lambda}=\tilde{\lambda}_r+i\tilde{\lambda_i}$, $\tilde{\lambda}_r>0$. 
Here $\sigma$ is a small parameter that controls the position of the initial condition on the plane spanned by the unstable eigenvectors.
Computing one thousand outward spiraling orbits with $\sigma=0.001\times\exp[(2\pi\tilde{\lambda}_r)/(j\tilde{\lambda}_i)]$, $j=1...1000$ then
results in an approximation of the unstable manifold. The stable manifold was computed similarly, sprinkling initial conditions along
the stable eigendirection and integrating backward in $\xi$.

The stable and unstable manifolds are best visualized by keeping track of their intersection with a \emph{Poincar\'e (surface of) section} $\mathcal{P}$, a $3$-dimensional surface in our $4$-dimensional phase space. Here we will be interested in symmetric homoclinic connections, which always intersect the symmetric section $\mathcal{S}$. It will therefore be important to choose a Poincar\'e section that contains $\mathcal{S}$, for instance $\dot{\phi}=0$.

\begin{figure}[h]
\begin{center}
\includegraphics[scale=0.4]{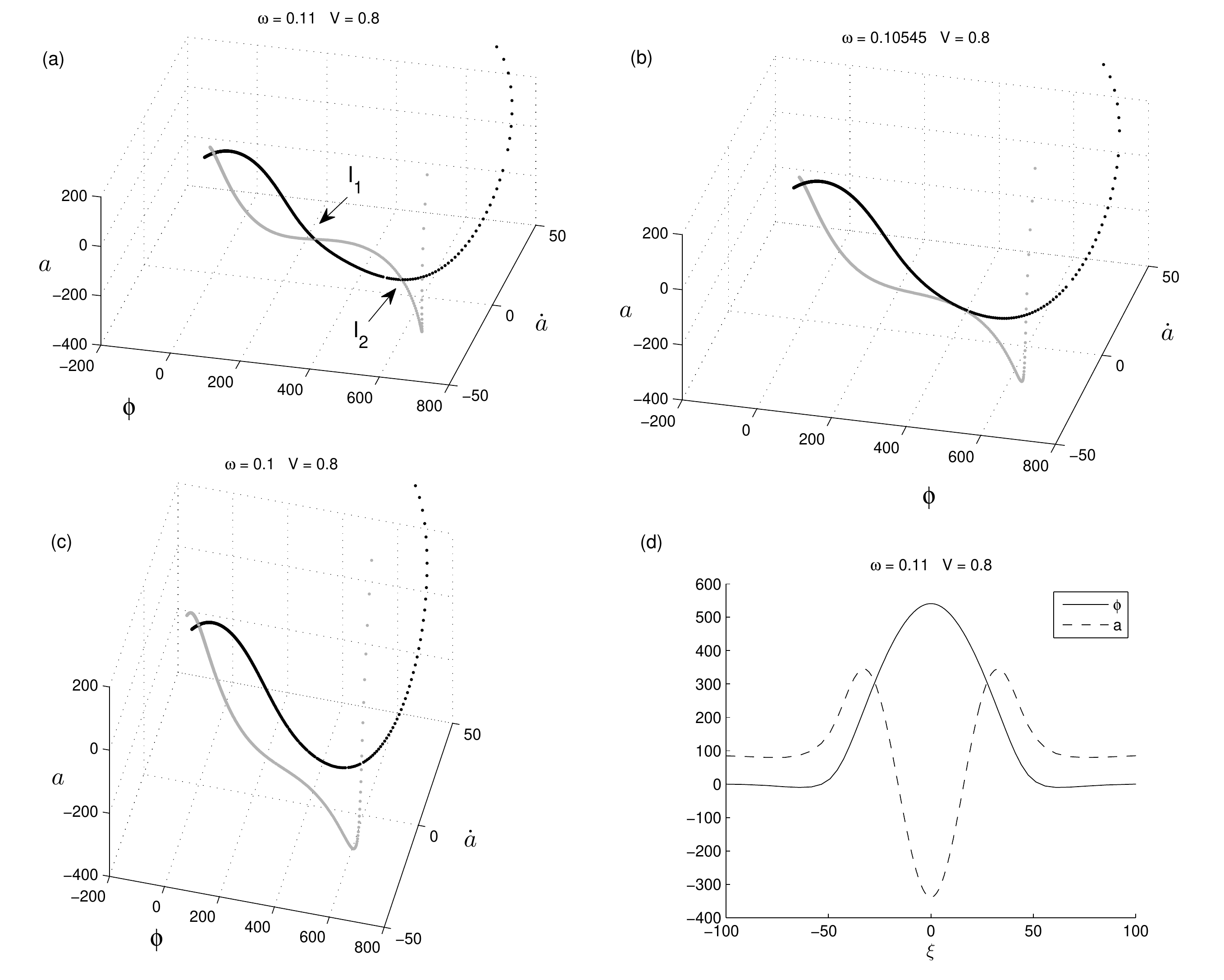}
 \caption{\label{Fig:Intersection} Panels (a)-(c) show the intersection of the stable (black) and the unstable (grey) manifolds with the Poincar\'e section $\dot{\phi}=0$. Panel (d) is the grey soliton corresponding to the point labelled $I_2$. }
\end{center}
\end{figure}

Figure \ref{Fig:Intersection} shows the first four intersections of our numerical approximation to the stable and unstable manifolds of $Q_0^+$ with $\mathcal{P}$, for three different values of $\omega$ with fixed $V=0.8$. For frequency value $\omega=0.11$ (panel \ref{Fig:Intersection}a) the stable and unstable manifolds intersect transversely and two homoclinic orbits (grey solitons) 
exist, marked by the points of intersection $I_1$ (belonging to the homoclinic orbit in figure \ref{Fig:HOMO1}(d)- \ref{Fig:HOMO1}(f)) and $I_2$ (belonging to the orbit  in figure  \ref{Fig:Intersection}(d)). At  $\omega\simeq0.10545$, $I_1$ and $I_2$ coalesce into a single solution and the manifolds become tangent. For $\omega\lesssim 0.10545$ the unstable and stable manifolds do not intersect and the family of homoclinic orbits shown in figure \ref{Fig:HOMO1} ceases to exist for $V=0.8$. Note however that different families of solitary waves do exist for $\omega\lesssim 0.10545$, $V=0.8$; in figure \ref{Fig:Intersection} we only plot a part of the stable and unstable manifolds and therefore more intersections can still take place. This becomes apparent when comparing, for instance, figures \ref{Fig:HOMO1} and \ref{Fig:HOMO2} showing that a family of grey solitons can extent beyond the range of existence of another family.

A similar analysis was carried out for the second type of grey solitons, figure \ref{Fig:HOMO2}. We computed the stable and unstable manifolds for parameter values close to the the boundary exhibited in figure \ref{Fig:HOMO2}(a) and we did not observed a tangency close to the homoclinic orbit, confirming that the existence boundary is a numerical artifact. Therefore, the Poincar\'e analysis is a useful tool to  distinguish real boundaries from numerical artifacts.  Note also that visualization of the stable and ustable manifolds of the fixed points on a Poincar\'e surface of section is a very effective way of generating initial guesses for the numerical computation of homoclinic and heteroclinic connections through the spectral method of section \ref{Sec:Algorithm}. 

\section{Stability of the solitary waves\label{Sec:Stability} }

Even though an exhaustive analysis of the stability of the solitary waves is beyond the scope of the present work, we sketch out here some relevant aspects. As previously mentioned, the  electromagnetic circularly polarized wave is susceptible to the relativistic Raman and the modulational instabilities that would ultimately destroy the solitary wave. However, since the growth rates of these instabilities are controlled by the amplitude of the electromagnetic wave ($a_0$ in our dynamical system) and the plasma density (related to the parameter $\omega$), certain parameter values could allow long distance propagation of the solitons. 

To test the stability of the solitary waves we show a couple of  1-dimensional simulations with the PIC code Calder \cite{Lefebvre03}. This method does not prove stability  of the solutions in a strict mathematical sense but it provides an insight into the dynamics. We took a computational domain equal to  $60\times 2\pi\sqrt{1-V^2}/\omega V$ (in $c/\omega_{pe}$ units) that is large enough to assume periodic boundary conditions and we used one million of cells with ten particles per cell. The code was  initialized with a grey solitary wave of the type presented in figure \ref{Fig:HOMO2}.

\begin{figure}[h]
\begin{center}
\includegraphics[scale=0.35]{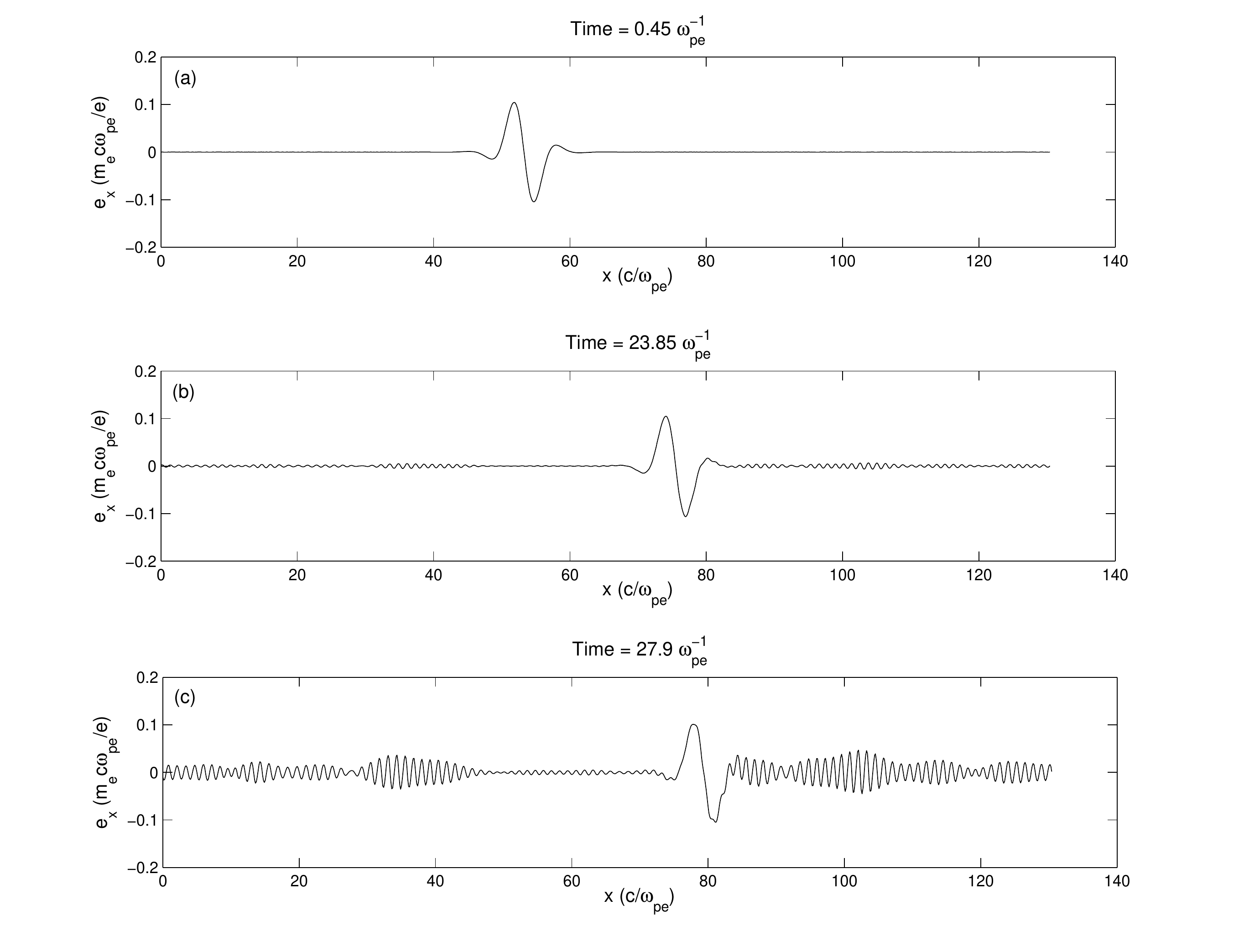}
 \caption{\label{Fig:Stability1}PIC simulation initialized with a grey soliton of the type shown in figure \ref{Fig:HOMO2} with $V=\omega = 0.95$.  }
\end{center}
\end{figure}

\begin{figure}[h]
\begin{center}
\includegraphics[scale=0.35]{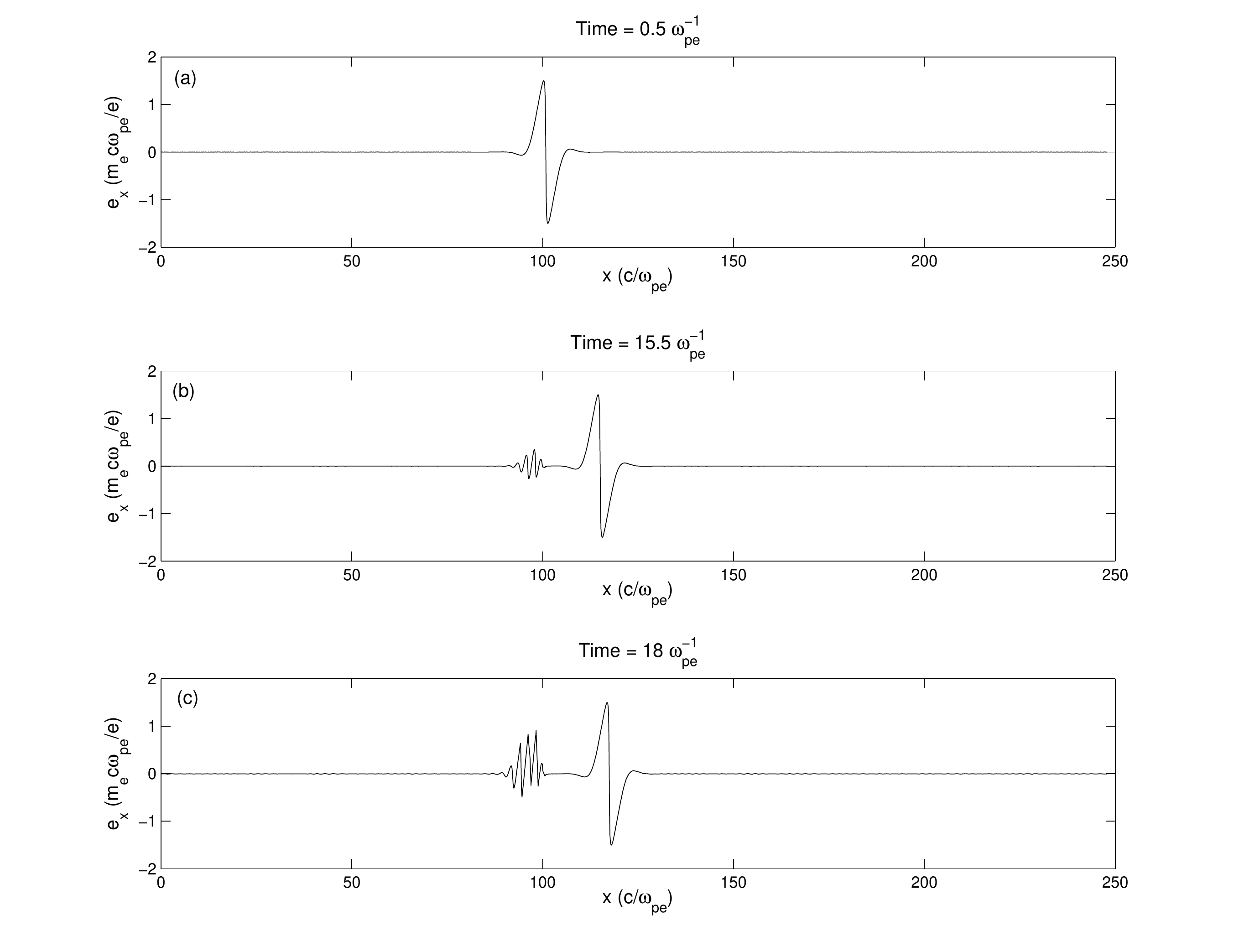}
 \caption{\label{Fig:Stability2}PIC simulation initialized with a grey soliton of the type shown in figure \ref{Fig:HOMO2} with $V= 0.95$ and $\omega = 0.5$. }
\end{center}
\end{figure}

 Figure \ref{Fig:Stability1}  shows the evolution of the electric field component $E_x$ of  a grey solitary wave with  $V = 0.95$ and  $\omega=0.95$. For this high value of the frequency the ions are almost immobile and the asymptotic value of the amplitude of the wave is $a_0\sim 0.48$. It propagates undistorted during a few tens of $\omega_{pe}^{-1}$, until the part of the solution with vanishing $E_x$ (corresponding to the infinitely long circularly polarized wave) becomes unstable due to the Raman instability. However, a solitary wave with a lower frequency  value $\omega=0.5$ ($a_0=3.88$) presents  different dynamics (see  figure \ref{Fig:Stability2}). For this second case the solitary wave develops an instability at its trailing edge and it radiates part of its energy away. A similar instability appears in multi-hump solutions with VBC \cite{Saxena06}.  These  examples reveal that, without collision and for a cold plasma model, our solutions present an  unstable character. However, a dissipation mechanism could reduce or even supppress the Raman instability, thus allowing the propagation of the solitary waves for longer distances.

\section{Conclusions}\label{Sec:Conclusions}

Solitary waves excited by  the interaction of a high-intensity laser with a plasma have been observed in laboratory experiments \cite{Borghesi02a,Borghesi02b,Pirozhkov07,Kando09,Sarri10,Romagnani10} and particle-in-cell (PIC) simulations \cite{Sentoku99,Bulanov99,Bulanov01,Esirkepov02}. Although many theoretical works have been carried out on these structures (see  \cite{Farina05} for a review), the parametric domain where the fixed point that controls the NVBC is a saddle-focus was unexplored. In our study of this regime we were able to exhibit  new ranges of solutions including grey and dark solitons. 

We recall that, for VBC, the solutions are organized in the $\omega-V$ plane on a set of infinitely  many branches \cite{Farina01a}. Each branch is characterized by the number of humps (or number of nodes) of the vector potential and it ends at a certain point due to the wavebreaking of the soliton. The potential is always a symmetric function whereas the vector potential can be either symmetric or antisymmetric. 

For NVBC we have shown that there is a continuum of solutions in the $\omega-V$ plane. Grey solitary waves with a symmetric potential and vector potential and dark solitary waves with symmetric potential and antisymmetric vector potential can be
found for a wide range of parameters. This is a natural extension of the  VBC case with the number of nodes being even or odd, respectively. Multi-hump solutions are also possible for any value of parameters for which a single-hump orbit exists. 
Further, asymmetric single and multi-hump solutions exist, consistent with symmetry breaking in a conservative, time-reversible system.

In addition to being an important  channel of laser-pulse energy transformation, solitary waves have been propossed as interesting candidates for photon and particle acceleration schemes \cite{Kaw92,Farina01a}. PIC simulations with an overcritical amplitude soliton showed electron acceleration during the nonlinear wavebreaking \cite{Esirkepov98} whereas ion acceleration has been detected during the postsoliton expansion \cite{Naumova01b}.  In  \cite{Farina01a}, the authors reported   wavebreaking of the solitary wave at the critical velocity determined by the end of the branch of the solutions in the $\omega-V$ plane.  It was estimated that ions could reach an energy value of the order of 70 MeV. Similarly, in the present work we have presented domains of existence in the $\omega-V$ plane and we have shown that coalescence of solitary waves (in parameter space) leads to disappeance of families of solutions. Coalescence was visualized by keeping track of the stable and unstable manifolds of a fixed point by means of a Poincar\'e surface of section. 

We point out that the stability of solitons with VBC has been studied in the past \cite{Naumova01b,Hadzievski02,Poornakala02,Poornakala02b,Lontano03,Lehmann06,Mancic06,Saxena07,Lehmann08}.  In particular, 1-dimensional  numerical fluid simulations with immobile ions showed that single hump solutions are stable whereas the multi-hump solutions suffer the Raman instability \cite{Saxena06,Saxena07}. On the other hand, 2-dimensional simulations revealed that all solutions are unstable and the tranverse dynamics always dominates the longitudinal one \cite{Lehmann08}. Our PIC simulations initialized with a grey solitary wave showed that some of them could propagate undistorted during a few tens of $\omega_{pe}^{-1}$, just before the circularly polarized wave suffers the Raman instability. However, other grey waves radiate a portion of their energy from the trailing edge, similarly to the multi-hump solutions with VBC \cite{Saxena06}. Since these are just a few examples, a complete stability analysis would requiry the study of other types of solutions (dark waves, asymmetric, multi-humps etc) in the whole $\omega-V$ plane. Adding  warm plasma effects or a collision term would be relevant too. 

Besides existence and stability, the question about how to excite solitary waves with NVBC remains open. However, soliton-like electromagnetic modes with VBC have been observed during laser plasma interaction \cite{Bulanov92}. Our preliminary PIC simulations on solitary wave excitation show that the interaction between a solitary wave with VBC and a long circularly polarized laser pulse can produce a solitary wave with NVBC. These simulations will be presented in a future work.     

\ack
G. S\'anchez-Arriaga is supported by ANR under the GOSPEL project, grant reference ANR-08-BLAN-0072-03. We thank D. B\'enisti and L. Gremillet for helpful discussions.

\appendix
\setcounter{section}{1}
\section*{Appendix. The existence and stability of the fixed points \label{Sec:Appendix}}

Let us write   \eref{sys:Hamiltonian:a}-\eref{sys:Hamiltonian:b}  as $dx/d\xi=f(x)$ with  $x=[\phi\  \dot{\phi}\  a \ \dot{a}]$. Fixed points are given by $\dot{\phi}=\dot{a}=0$ and the solutions of

\begin{eqnarray}
\left[V\left(\frac{1}{r_e}+\frac{\epsilon}{r_i}\right)-\omega^2\right]a&=&0\label{Eq:FP1}\\
V\left(\frac{\psi_e}{r_e}-\frac{\psi_i}{r_i}\right)&=&0\label{Eq:FP2}
\end{eqnarray}
whereas the stability depends on the eigenvalue of the Jacobian matrix
\footnotesize
\begin{equation}
\fl
J=\left(\begin{array}{c c c c}
0                                                                                                    & 1 &          0                                                                                           & 0 \\
-V(1-V^2)\left(\frac{1+a^2}{r_e^3}+\frac{\epsilon(1+\epsilon^2a^2)}{r_i^3}\right)                        & 0 & V(1-V^2) a\left(\frac{\Gamma_{e}+\phi}{r_e^3}-\frac{\epsilon_i^2(\Gamma_{i}-\epsilon\phi)}{r_i^3}\right) & 0 \\
0                                                                                                    & 0 &          0                                                                                           & 1 \\
 -Va\left(\frac{\Gamma_{e}+\phi}{r_e^3}-\frac{\epsilon^2(\Gamma_{i}-\epsilon\phi)}{r_i^3}\right)& 0 &-\omega^2+V\left(\frac{1}{r_e}+\frac{\epsilon}{r_i}\right)+V(1-V^2) a^2\left(\frac{1}{r_e^3}+\frac{\epsilon^3}{r_i^3}\right) &0
\end{array}\right)
\end{equation}
\normalsize

The four eigenvalues of $J$ at the fixed points $Q_0^{\pm}$ can be written as $\lambda^2_{1-4}=-\delta\pm \sqrt{\Delta}$ with
\begin{equation}
\fl
\delta(V,a_0) \equiv \frac{1-V^2}{2V^2}\left(\frac{1}{\Gamma_{e}^3}+\frac{\epsilon}{\Gamma_{i}^3}\right)>0
\end{equation}
\begin{equation}
\fl
\Delta(V,a_0) \equiv \left[\frac{1-V^2}{2V^2}\left(\frac{\Gamma_{e}^2+a_0^2}{\Gamma_{e}^3}+\epsilon\frac{\Gamma_{i}^2+\epsilon^2a_0^2}{\Gamma_{i}^3}\right)\right]^2 -\frac{(1-V^2)(1-\epsilon^2)^2a_0^2}{(V\Gamma_{e}\Gamma_{i})^4}
\end{equation}
Fixed point  $\displaystyle{Q_0^{+}}$ (or $\displaystyle{Q_0^{-}}$ ) is a saddle-focus if $\displaystyle{\Delta<0}$, a saddle-center if  $\displaystyle{\sqrt{\Delta}>\delta>0}$ and a center if  $\displaystyle{0<\sqrt{\Delta}<\delta}$. The conditions $\sqrt{\Delta}=\delta$ and  $\Delta=0$ yield the velocities \eref{Eq:Vsc} and \eref{Eq:Vsf} respectively.

The eigenvalue of the Jacobian matrix at $Q_1$ for $V=V_s$ are
\begin{equation}
\lambda_{1,2}=\pm\omega\sqrt{\frac{(\Gamma_{e}\Gamma_{i}-1)(\Gamma_{e}-\Gamma_{i})^2}{2(\Gamma_{e}-\Gamma_{i})^2+(\Gamma_{i}+\epsilon\Gamma_{e})^2a_0^2}}
\end{equation}
\begin{equation}
\lambda_{3,4}=\pm i(1+\epsilon)^2\sqrt{\frac{V_s(1-V_s^2)}{\left[(\Gamma_{i}+\epsilon_i\Gamma_{e})^2-(1-V_s^2)(1+\epsilon)^2\right]^{3/2}}}
\end{equation}
and clearly the fixed point $Q_1$ is a saddle-center (note that $\Gamma_e,\Gamma_i>1$).  

On the other hand, the existence and stability analysis of $Q_2$ requires some auxiliar operations. From  \eref{Eq:FP1} and assuming $-\Gamma_e<\phi_f<\Gamma_i/\epsilon$, one gets
\begin{equation}
 a_f^2=\frac{\psi_e^2-\psi_i^2}{\psi_i^2-\epsilon^2\psi_e^2}\label{Eq:Q3:a}
\end{equation}
and by substituting  in  \eref{Eq:FP2}.
\begin{equation}
G(\phi_f)\equiv \frac{1}{\Gamma_{e}\Gamma_{i}}-\frac{V}{\psi_e\psi_i}\sqrt{\frac{\psi_i^2-\epsilon^2\psi_e^2}{\psi_i^2-\epsilon^2\psi_e^2-(1-V^2)(1-\epsilon^2)}}=0\label{Eq:G}
\end{equation}
where we introduced the subscript $f$ to denote that we are dealing with a fixed point. Note that restriction \eref{Eq:bound:general} together with $a^2>0$ in \eref{Eq:Q3:a} show that solutions of  \eref{Eq:G} must lie on the intervals

\begin{equation}
 \phi_{min2}\equiv-\frac{\Gamma_{e}-\Gamma_{i}}{1+\epsilon}<\phi_f<\frac{(1-\epsilon^2)V^2}{2\epsilon(\Gamma_{i}+\epsilon\Gamma_{e})}\equiv \phi_{max2}
\end{equation}

The solutions of \eref{Eq:G} can be discussed taking into account some properties of the function $G(\phi_f)$ and its derivative: 
\footnotesize
\begin{equation}
\fl
\frac{dG}{d\phi_f}=\frac{V}{\psi_e\psi_i}\sqrt{\frac{\psi_i^2-\epsilon^2\psi_e^2}{\psi_i^2-\epsilon^2\psi_e^2-(1-V^2)(1-\epsilon^2)}} \left[\frac{\psi_i-\epsilon\psi_e}{\psi_e\psi_i}-\frac{\epsilon(1-V^2)(1-\epsilon^2)}{\left[\psi_i^2-\epsilon^2\psi_e^2-(1-V^2)(1-\epsilon^2)\right](\psi_i-\epsilon\psi_e)}\right]
\end{equation}
\normalsize
In particular we are interested in the zeros of this derivative. The factor inside the squared root vanishes at $\phi_f=\phi_{max2}/V^2>\phi_{max2}$, that it is outside the physical domain. On the other hand, the roots of the term inside the braces are given by the zeros of the cubic equation 
\footnotesize
\begin{equation}
\fl
\frac{8(\Gamma_i+\epsilon\Gamma_e)}{1-\epsilon^2}\epsilon^3\phi_f^3-3(3+V^2)\epsilon^2\phi_f^2+3(\Gamma_i-\epsilon\Gamma_e)(1+V^2)\epsilon\phi_f-(\Gamma_i-\epsilon\Gamma_e)^2V^2+\epsilon(1-V^2)\Gamma_e\Gamma_i=0
\ \label{Eq:cubic}
\end{equation}
\normalsize

For discussing the solutions of  \eref{Eq:G} within the domain  $\phi_{min2}<\phi_f<\phi_{max2}$, we first note that $G(0)=0$,  corresponding with the fixed points $Q_0^{\pm}$. One also has  the asymptotic behaviours $G\rightarrow -\infty$ as $\phi_f\rightarrow \phi_{max2}$ and $G(\phi_{min2})>0$ ($<0$) for velocities less (greater) than $v$ (see  \eref{Eq:V:Q2}). The analysis of its derivative shows that $G$ has 0 or 1 extreme for $V<v$ and therefore  \eref{Eq:G} has one solution in this regime ($Q_0^{\pm}$). On the other hand, for  $V>v$ it always has one extreme, at say $\phi_f^*$, with $G(\phi_f^*)\ge 0$. Therefore, if $G(\phi_f^*)> 0$,  \eref{Eq:G} has two solutions and, in addition to   $Q_0^{\pm}$, there is another fixed point that we call $Q_2$. One also checks that  $Q_2$ and  $Q_0^{\pm}$ lie in different manifolds given by  \eref{Eq:Inv}. Hence, heteroclinic connections among $Q_2$ and  $Q_0^{\pm}$ are not possible.

\section*{References}

\providecommand{\newblock}{}

\end{document}